\documentclass[%
 reprint,
superscriptaddress,
 amsmath,amssymb,
 aps,
]{revtex4-2}

\usepackage{graphicx}
\usepackage{dcolumn}
\usepackage{bm}
\usepackage{xcolor}

\begin{document}

\preprint{APS/123-QED}

\title{Epitaxial stabilization of magnetic GdAuSb/LaAuSb superlattices}

\author{Patrick J. Strohbeen}
    \affiliation{Department of Materials Science, University of Wisconsin Madison, Madison, Wisconsin 53706, USA}

\author{Soohyun Im}
    \affiliation{Department of Materials Design and Innovation, The State University of New York, Buffalo, New York 14260, USA}

\author{Tamalika Samanta}
    \affiliation{Department of Materials Science, University of Wisconsin Madison, Madison, Wisconsin 53706, USA}

\author{Zachary LaDuca}
    \affiliation{Department of Materials Science, University of Wisconsin Madison, Madison, Wisconsin 53706, USA}
    
\author{Dongxue Du}%
    \affiliation{Department of Materials Science, University of Wisconsin Madison, Madison, Wisconsin 53706, USA}
    
\author{Estiaque H. Shourov}
    \affiliation{Department of Materials Science, University of Wisconsin Madison, Madison, Wisconsin 53706, USA}%
    
\author{Jessica L. McChesney}
    \affiliation{Argonne National Laboratory, 9700 South Cass Avenue, Argonne, Illinois 60439, USA}

\author{Fanny Rodolakis}
    \affiliation{Argonne National Laboratory, 9700 South Cass Avenue, Argonne, Illinois 60439, USA}

\author{Paul M. Voyles}
    \affiliation{Department of Materials Science, University of Wisconsin Madison, Madison, Wisconsin 53706, USA}

\author{Jason K. Kawasaki}
    \thanks{Corresponding Author: jkawasaki@wisc.edu}
    \affiliation{Department of Materials Science, University of Wisconsin Madison, Madison, Wisconsin 53706, USA}

\date{\today}

\begin{abstract}
We report the epitaxial stabilization of GdAuSb films and GdAuSb/LaAuSb superlattices via molecular beam epitaxy on (0001)-oriented Al$_{2}$O$_{3}$ substrates. GdAuSb crystallize in the Au-Au dimerized YPtAs structure type (space group $P6_{3}/mmc$), the same structure as the Dirac semimetal LaAuSb. Angle-resolved photoemission spectroscopy (ARPES) measurements show similar near $E_F$ bandstructures for GdAuSb and LaAuSb, plus a rigid band shift for GdAuSb towards more hole-like behavior and core-like Gd $4f$ states $\sim 9$~eV below the Fermi energy. LaAuSb/GdAuSb superlattices exhibit sharp superlattice fringes by X-ray diffraction and atomically-precise interfaces by scanning transmission electron microscopy. Superlattices display two transitions in temperature-dependent resistvity, compared to a single N\'{e}el temperature for thick GdAuSb films. Superlattices of $Ln$AuSb materials ($Ln=$ rare earth) with atomically abrupt interfaces offer a new epitaxial platform for control of magnetic and topological order via tunable intralayer exchange and reduced dimensionality.
\end{abstract}

\maketitle

Rare earth-containing hexagonal \textit{ABC} ($A = Ln$) compounds offer opportunities for tunable intrinsic polar metallicity~\cite{du2019laglps, du2024polardist}, Weyl semimetal-like behavior~\cite{du2023gdptsbvac}, and strain gradient-induced magnetism (flexomagnetism)~\cite{laduca2024flexo}. The most well-studied compounds within this class of materials crystallize in a closed-shell, 18-valence electron configuration (e.g. $Ln$AuGe, $Ln$PtSb, LiGaGe-type structure, Polar space group $P6_{3}mc$). Lanthanide ($A$-site) substitutions are a powerful tool for controlling and tuning the magnetic behavior within these systems for realizing engineereable Weyl semimetals~\cite{manna2018hwb, duan2018repntuning, ho2023repntopology, kushnirenko2022repnfermi} within this family. Polar distortions aredictated by the observed buckling in the $BC$-plane, which is tunable via strain and chemical pressure~\cite{ddu2019polarmetals, du2024polardist}.

Adjacent to the 18-valence electron compounds, the family of $Ln$AuSb compounds has similar chemical composition to the 18-valence counterparts, but with one extra valence electron per formula unit. These 19-valence electron compounds have also been shown recently to be of interest as a candidate bulk topological semimetal~\cite{seibel2015gold}. The electronic instability caused by the excess valence electron is stabilized through the formation of a Au-Au dimer bond along the $c$-axis of the crystal structure, resulting in a comparative doubling of the unit cell along this axis and causing the crystal to become non-polar~\cite{seibel2015gold, strohbeen2019LaAuSb}. In addition to the expected bulk topological band crossings~\cite{seibel2015gold}, partially filled $4f$ bands via lanthanide substitutions further allow for the exploration of the interplay of tunable magnetism and non-trivial band topology similar to what has been done previously in their 18-valence electron counterparts. Furthermore, previous studies on single crystals of the LaAuSb compound~\cite{strohbeen2019LaAuSb} found that the Au-Sb buckling value is drastically enhanced compared to the non-centrosymmetric 18-valence analogues~\cite{ddu2019polarmetals}, owing to the presence of these Au-Au dimers. 

However, while the 18-valence count compounds is well-studied and has demonstrated the ability for alloying across the full range of the Lanthanide species, studies on the 19-valence variant are notably less mature. Preliminary bulk studies in the 19-valence electron crystals have observed limitations in allowed alloy formation to only the early lanthanide species (Ln = La-Nd, Sm; or $f$ occupancies of 0-3, 5)~\cite{seibel2015gold}. Furthermore, limitations imposed by the centrosymmetric nature of the 19-valence electron compounds must be overcome in order to utilize the enhanced layer buckling caused by the Au-Au dimerization, i.e. robust methods for breaking inversion symmetry in these compounds need to be investigated. Therefore, it is of interest to develop thin film growth of 19-valence electron $Ln$AuSb ternary intermetallic thin films to enable approaches for engineering crystal symmetry and emergent magnetic properties within these compounds. 

\begin{figure*}
    \centering
    \includegraphics[width=\linewidth]{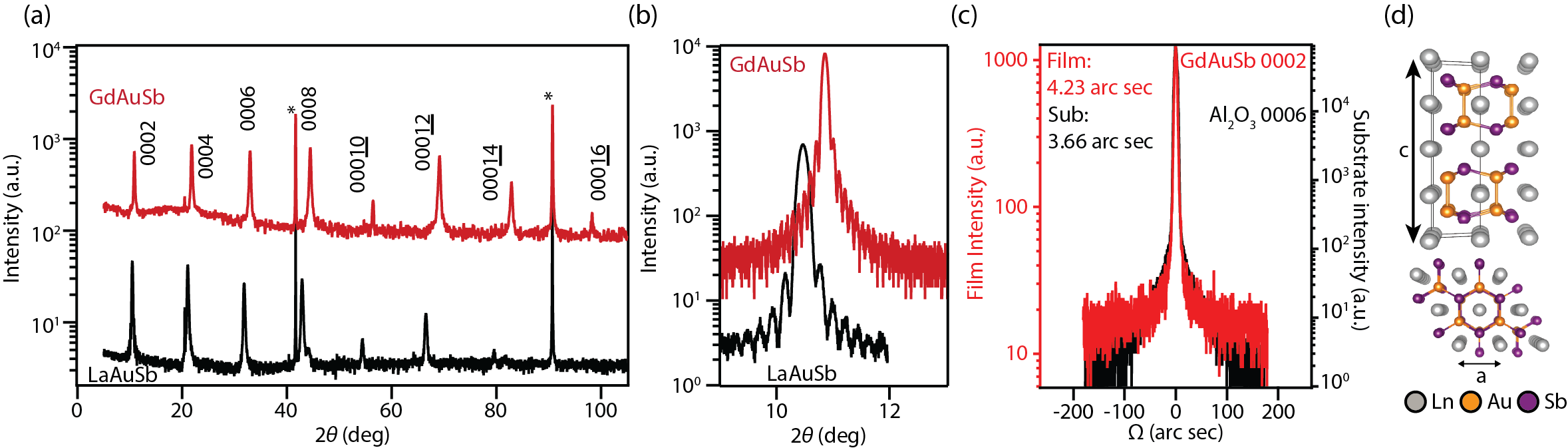}
    \caption{\textbf{(a)} Survey $\omega$-2$\theta$ x-ray diffraction scan comparing the new GdAuSb compound to the previously reported LaAuSb alloy. 000\textit{l} reflections are labeled, substrate reflections are labeled with asterisks (*). \textbf{(b)} Zoom-in on (0002) reflection for GdAuSb and LaAuSb films to highlight the pronounced Keissig fringes. \textbf{(c)} Rocking curve about the GdAuSb (0002) and Al$_{2}$O$_{3}$ (0006) reflections displaying the low mosaicity exhibited in the film. \textbf{(d)} Schematic crystal structures of the 19-valence \textit{Ln}AuSb compounds crystallizing in the YPtAs structure.}
    \label{xtal_struc}
\end{figure*}

Here, we present epitaxial stabilization of GdAuSb, a hexagonal \textit{ABC} Heusler alloy, which we force to crystallize in the YPtAs-type (Space group No. 194, $P6_{3}/mmc$) structure that does not exist in bulk form~\cite{seibel2015gold}. Using angle-resolved photoemission spectroscopy (ARPES) we show that LaAuSb and GdAuSb display a similar near $E_F$ electronic structure, plus a rigid band shift for GdAuSb and a partially occupied \textit{f}-band 9~eV below E$_{F}$. Lastly, we present this hexagonal intermetallic family of \textit{Ln}AuSb as an intriguing platform for study of intermetallic alloy superlattices. We find sharp superlattice interfaces compared to previous Heusler superlattice attempts of the $L2_{1}$ phase in the literature~\cite{brownheft2018heuslersl} without needing to drastically reduce crystalline quality via low-temperature nucleation. Superlattices of GdAuSb/LaAuSb display two transitions in resistivity vs temperature, compared to the single N\'{e}el transition for thick GdAuSb films. The thin film epitaxy and superlattice approach demonstrated here shows the promise of these two routes towards engineering magnetism and non-trivial band topology within the family of 19-valence $Ln$AuSb materials. 

We first demonstrate the epitaxial synthesis of GdAuSb and LaAuSb on Al$_2$O$_3$ (0001) substrates. LaAuSb thin films and bulk crystals have been previously reported in Refs.~\cite{strohbeen2019LaAuSb} and~\cite{seibel2015gold}, respectively. GdAuSb has not previously been synthesized in bulk crystal or thin film form ~\cite{seibel2015gold}. Figure~\ref{xtal_struc}a shows the $\omega-2\theta$ scan of GdAuSb, compared to data for  LaAuSb ~\cite{strohbeen2019LaAuSb}. For GdAuSb we observe only the expected $000l$ reflections and no secondary phases indicative of a singly-oriented film in the growth direction. The presence of the 0002, 0006, 000\underline{10}, 000\underline{14}, and 000\underline{16} superstructure reflections in the GdAuSb confirms the doubling of the unit cell along the $c$ axis, consistent with the Au-Au dimerization observed for LaAuSb~\cite{strohbeen2019LaAuSb, seibel2015gold} (Figure~\ref{xtal_struc}d). The Keissig fringe spacing around the $0002$ reflection, shown in Figure~\ref{xtal_struc}b, corresponds to a film thickness of 85.5nm, consistent with our flux measurements and RHEED oscillations. From Figure~\ref{xtal_struc}a, we calculate the out-of-plane $c$-axis parameter of GdAuSb to be 16.33~\r{A}, slighly larger than the expected value of 16.18~\r{A} but in reasonably good agreement. We calculate the expected lattice parameter based on a linear extrapolation of the lanthanide contraction following the trend from LaAuSb, CeAuSb, and NdAuSb bulk crystal structures~\cite{seibel2015gold}. Figure~\ref{xtal_struc}c presents the rocking curve data for GdAuSb overlaid with that of the underlying c-plane sapphire substrate. Rocking curve measurements for the film and substrate display sharp reflections and similar line widths of 4.23 and 3.66 arc seconds, respectively, indicating a minimal mosaic spread in the film.

\begin{figure*}
    \centering
    \includegraphics[width=\linewidth]{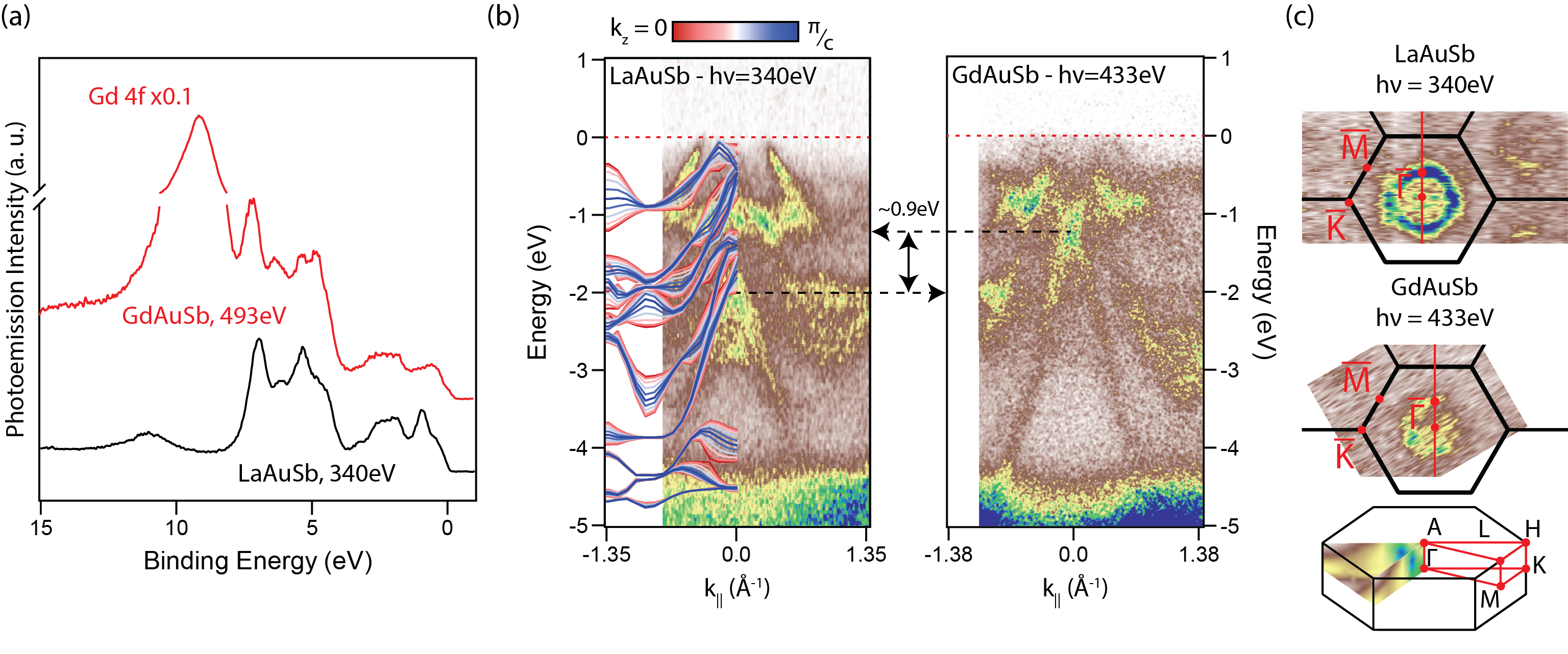}
    \caption{\textbf{Electronic structure of thick GdAuSb and LaAuSb films.} \textbf{(a)} Angle-integrated photoemission spectroscopy of the valence bands for GdAuSb and LaAuSb. The Gd 4f states are on a split scale at 0.1x the original scale to fit the states into the same plot.
    \textbf{(b)} Angle-resolved photoemission spectroscopy (ARPES) for LaAuSb (left) and GdAuSb (right) taken along $\overline{\Gamma}-\overline{M}$ directions, as indicated by the red line in (c). The overlaid traces are simulated band structure calculated using the WIEN2K DFT package~\cite{wien2k}. The colors refer to specific $k_z$ values ranging from 0 ($\Gamma$, red) to 0.385~\r{A}$^{-1}$ ($A$, blue) in equal steps of 0.064~\r{A}$^{-1}$
    \textbf{(c)} Fermi surface for LaAuSb and GdAuSb with schematic projected surface Brillouin Zone with high symmetry points $\overline{\Gamma}, \overline{M},$ and $\overline{K}$ labeled. The red line represents the cut taken for the dispersion presented in \textbf{b}. The 3-Dimensional Brillouin Zone is also presented with the high symmetry points of the hexagonal system labeled. The color plots are DFT calculations for the Fermi surface showing the minimal dispersion along the $\Gamma$-A direction.}
    \label{elec_struc}
\end{figure*}

Figure~\ref{elec_struc}a presents the angle integrated photoemission of the valence bands for GdAuSb and LaAuSb, showing the highly similar Near-$E_\textup{F}$ density of states for the two alloys. GdAuSb displays an additional peak at $\sim 9$~eV corresponding to Gd $4f$-states~\cite{lbldatabooklet}. Due to their relatively large binding energy, we expect the $f$-orbital states to behave as core-like, i.e. we do not expect significant hybridiztion of the $4f$-states with the near $E_\textup{F}$ bands in GdAuSb. The dispersion along the $\overline{\Gamma}-\overline{M}$ path (see Figure~\ref{elec_struc}c for the 2D projected and 3D Brillouin zones) is presented in Figure~\ref{elec_struc}b along with an overlaid bandstructure calculation for varying values along $k_\textup{z}$ (0 - $\pi$/c). The calculation clearly demonstrates the quasi-2D nature of the LaAuSb bandstructure, as shown by the minimal dispersion along $k_\textup{z}$. This is further emphasized in our experiments due to the large broadening in $k_\textup{z}$ ($\Delta k_\textup{z} \sim 1/ \lambda_{imfp}$) when measuring in the tender X-ray regime. We calculate an expected broadening in $k_\textup{z}$ to be 0.1~\r{A}$^{-1}$ and 0.09~\r{A}$^{-1}$ for photons with kinetic energies of 340~eV and 433~eV, respectively~\cite{tanuma2003calculation}, or roughly 0.27$\pi$/c for LaAuSb. Further demonstration of the cylindrical nature of the Fermi surface is shown in Figure~\ref{elec_struc}c in the 3D Brillouin zone with the DFT calculated Fermi surface overlaid. The resulting $k_\textup{z}$-broadened ARPES measurement displays excellent agreement with the broadened DFT calculation. However, there is a bright hole pocket crossing the Fermi level that is not captured at all by the DFT calculation. We attribute this bright feature to a surface state that is not captured by the bulk DFT calculation, but we were unable to experimentally confirm this due to the significant $k_z$ broadening and pseudo-2D nature making it difficult to measure dispersion along $k_\textup{z}$. 

Comparing the angle-resolved bandstructure along the $\overline{\Gamma} - \overline{M}$ cut, as indicated in Figure~\ref{elec_struc}b with the black dashed lines serving as guides to the eye, we observe qualitatively very similar electronic structure between the Gd-containing and La-containing compounds. Notably, we observe evidence of the expected topologically non-trivial band crossing in LaAuSb~\cite{seibel2015gold} near -2~eV, shifting up in energy in the Gd-containing compound by roughly 0.9~eV to around -1.1~eV. In other words, it appears that the major difference between these alloys is an increase in hole-like behavior in the Gd-containing compound with minimal perturbation otherwise to the zone center band dispersion.

We next synthesize GdAuSb/LaAuSb superlattices, towards the goal of tuning Gd-Gd interlayer exchange as a function of LaAuSb spacer thickness. The overall film structure is [(LaAuSb)$_{m}$/(GdAuSb)$_{n}$]$_{N}$/GdAuSb buffer/Al$_{2}$O$_{3}$, where the purpose of the thick ($\sim$43~nm), relaxed GdAuSb buffer layer is to obtain a smooth surface for before initiating superlattice growth. Here, $m$ and $n$ are the number of formula units of LaAuSb and GdAuSb, respectively, and $N$ is the number of superlattice repeat units. Superlattice growth is conducted by supplying constant Au and Sb fluxes at the same rate as mentioned earlier in the text while alternating the La and Gd shutters.

\begin{figure*}
    \includegraphics[width=\linewidth]{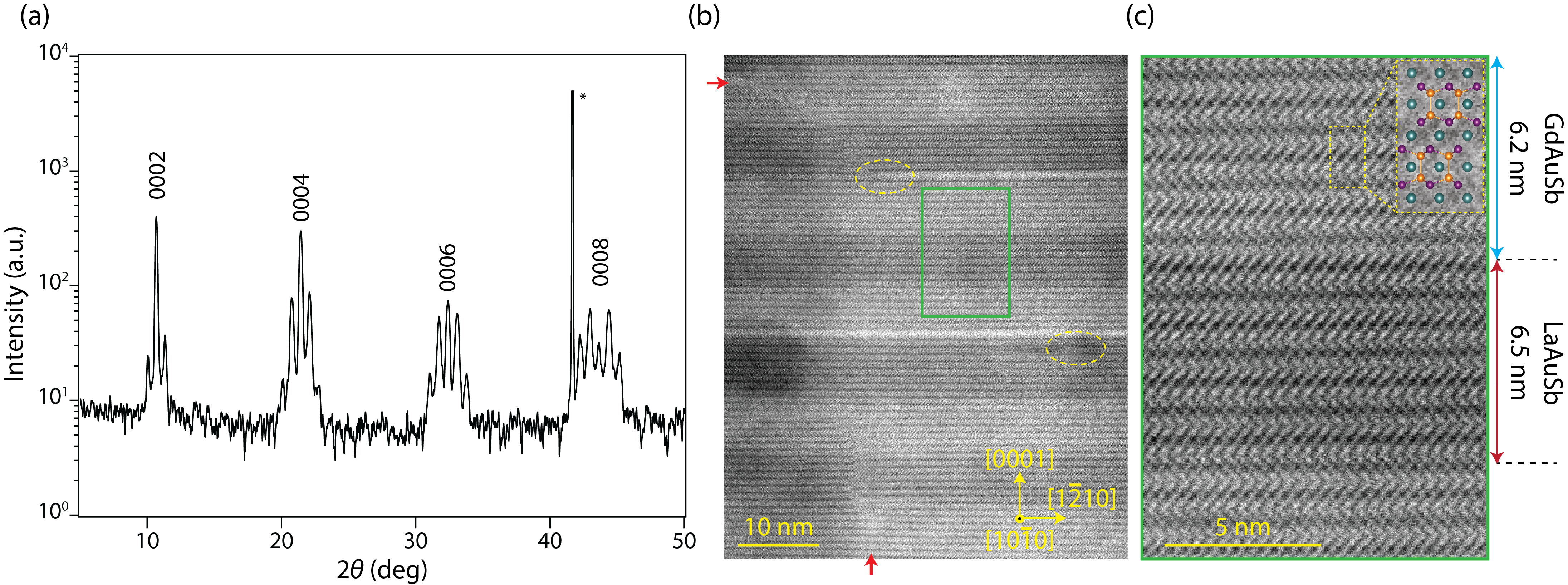}
    \caption{\textbf{(a)} Exemplary long-range $\theta$-$2\theta$ out-of-plane XRD scan of [(LaAuSb)$_{8}$/(GdAuSb)$_{8}$] superlattice structure grown on a sapphire (0001) substrate. \textbf{(b)} Cross-sectional STEM image taken along a $[10\overline{1}0]$ direction on the $[0001]$ zone axis of the same [(LaAuSb)$_{8}$/(GdAuSb)$_{8}$] superlattice structure in (a). The dotted yellow circles highlight stacking fault defects and the sharp horizontal white lines we believe to be due to Gd-rich off-stoichiometries. The dark and light stripes are the LaAuSb and GdAuSb layers, respectively. The red arrows denote a line defect that traces through the frame of the image. \textbf{(c)} Zoom in on region defined in \textbf{(b)}. The LaAuSb and GdAuSb layer thicknesses are presented and the inset shows a zoom-in on the GdAuSb crystal structure with an overlaid ball-and-stick model to signify the atoms. In the schematic, green atoms are Gd, purple atoms are Sb, and gold atoms are Au.}
    \label{SL_struct}
\end{figure*}

Structural characterization of an exemplary $m=n=8$, $N=5$ superlattice is presented in Figure~\ref{SL_struct}. Fig.~\ref{SL_struct}a shows a long-range XRD survey of the superlattice. The presence of the superlattice reflections up to second order suggest a high degree of interfacial stability in this system. From these fringes, we calculate the superlattice period to be 6.33~nm, which agrees reasonably well with the intended period thickness of 6.66~nm calculated from our growth fluxes. The sharp LaAuSb/GdAuSb interfaces are further supported by cross-sectional STEM imaging, seen in Fig.~\ref{SL_struct}b. Here the interface between the two layers remains sharp, which is typically difficult for intermetallic alloys at elevated growth temperatures~\cite{brownheft2018heuslersl, mancoff1999nimnsbc1bsl}.

Several extended defects are also present in the field of view shown in Fig.~\ref{SL_struct}b. We observe a dislocation line entering through the bottom of the frame and exiting near the top, as indicated by the red arrows. This dislocation gives rise to a stacking fault defect across the dislocation boundary and when this stacking fault exists at the GdAuSb/LaAuSb interface, appears to cause a build-up of excess Gd confined to a single layer. When the stacking fault does not exist at the interface between two superlattice layers, we instead do not observe any clear build-up of excess Gd or La atoms. We observe both features in Figure~\ref{SL_struct}b, as notated in the yellow dashed-line circles, where the top-most circle highlights a stacking fault at the GdAuSb/LaAuSb interface while the second yellow dashed-line circle displays a second stacking fault in an LaAuSb layer. Figure~\ref{SL_struct}c is a zoom in image of the region labeled by the green rectangle in Fig.~\ref{SL_struct}b to more clearly observe the crystal structure of the superlattice. The inset presents a ball-and-stick model of a single dimerized unit cell. We observe a robust layer structure and minimal chemical intermixing between the two compounds.

\begin{figure*}
    \centering
    \includegraphics[width=0.9\linewidth]{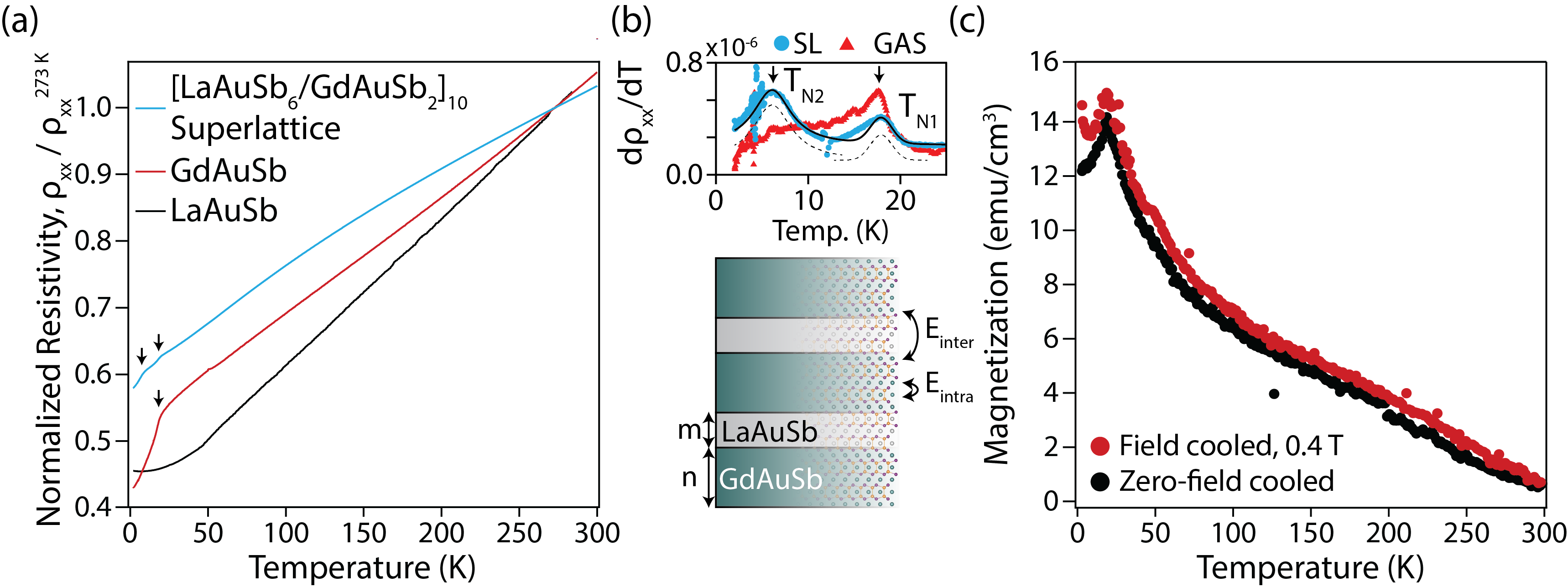}
    \caption{\textbf{a} Longitudinal resistivity versus temperature for GdAuSb, LaAuSb, and \textit{n}=2, \textit{m}=6 superlattice structure. The resistivity is normalized to the value at 273~K for all three samples to remove effects from interface scattering in the superlattice structure. We observe one transition at 17.85~K in both the pure GdAuSb and superlattice samples, and in the latter case a second transition is observed at 6.13~K. \textbf{b} Presents the derivative of the sheet resistance with respect to temperature for both the superlattice (blue circles) and the pure bulk-like GdAuSb film (red triangles). The transition temperatures were extracted by fitting the two peaks as shown by the black lines where the dashed lines are the individual peaks. We note an asymmetry in the peak structure in $T_\textup{N1}$ that appears in both samples is not present in the lower temperature transition, thus we attribute this to the thick GdAuSb buffer layer and not an effect caused by the superlattice. The schematic of the superlattice presents the different layer thicknesses and how changing LaAuSb spacer layer thickness affects the interlayer exchange coupling between GdAuSb layers.
    \textbf{c} Magnetization measurements of pure GdAuSb thin films in a zero-field cooled (black) and field cooled (red) configuration. Field cooling is conducted at 0.4~T, and the data was taken during warm-up. A clear antiferromagnetic transition is observed at 18~K, which agrees well with the transition $T_\textup{N1}$ observed in the longitudinal resistivity measurement.}
    \label{SL_transport}
\end{figure*}

Exemplary traces of normalized resistivity as a function of temperature during cooldown for GdAuSb, LaAuSb, and a $m=6, n=2$ superlattice are presented in Figure \ref{SL_transport}a. Whereas the resistivity of LaAuSb decreases monotonically with decreasing temperature, for GdAuSb we observe a kink corresponding to a N\'{e}el transition that we also observe by SQUID magnetometry (Fig. \ref{SL_transport}c). The $m=6, n=2$ superlattice exhibits the same resistivity kink as the pure GdAuSb film, plus a second transition that is shifted to lower temperature, as labeled in Figures~\ref{SL_transport}a and b by the black arrows. To clarify these transitions, we take the first derivative of the longitudinal resistivity with respect to temperature for both the superlattice and pure GdAuSb films and plot this in Figure~\ref{SL_transport}b. We fit the observed peaks to extract the temperatures of both N\'{e}el transitions, where the first N\'{e}el transition, T$_\textup{N1}$, occurs at 17.85~K, while the lower temperature transition occurs at T$_\textup{N2}=6.13$~K. The fits are shown in Figure~\ref{SL_transport}b. The black dashed lines are the individual peak components and the solid black line is the total fit. We note a multi-component lineshape for T$_\textup{N1}$ in the superlattice and bulk GdAuSb films, seen in the data as an asymmetry in the lineshape, that is not present in T$_\textup{N2}$. Since this feature is observed in the parent compound and not an effect caused by the superlattice, further investigation into the magnetic behavior of GdAuSb is needed, though it is out of scope for this superlattice study. We attribute the lower temperature transition, T$_\textup{N2}$, observed only in the superlattice sample, to the bulk-like Ne\'{e}l transition (T$_\textup{N1}$) that has been shifted down in temperature. We explain this behavior in the schematic shown in Figure~\ref{SL_transport}b. For GdAuSb, antiferromagnetic exchange coupling with energy $E_\textup{intra}$ induces the Ne\'{e}l transition at 17.85~K. However, when grown in a superlattice with non-magnetic spacer material (LaAuSb), a new exchange coupling is introduced, $E_\textup{inter}$, as the coupling between GdAuSb layers. As the spacer thickness increases $E_\textup{inter}$ decreases far enough to no longer induce a global antiferromagnetic phase until significantly lower temperatures.

To first order, if we consider the $4f$-states to exponentially decay following a typical plane wave behavior (i.e. $\Phi(R) \sim Exp[-i\vec{k}\cdot \vec{r}]$), a notable dropoff in exchange strength would be expected at a distance of 3 unit cells away, e.g. $Exp[-3] \sim 0.05$. We take this to be a lower limit on the expected distance between magnetic units to observe a difference in magnetic sructure of these systems. More explicitly, as seen in Figure~\ref{elec_struc}, the GdAuSb compound exhibits a quasi-cylindrical Fermi surface which we expect to display a slower fall off, assuming purely Ruderman-Kittel-Kasuya-Yosida (RKKY)-like antiferromagnetic exchange interactions that fall off following an inverse square law, i.e. $\Phi(R) \sim 1/\mathbf{R^{2}}$, ~\cite{roth1966rkky} in such systems. In our structure presented here, we have grown the sample such that the average spacing between GdAuSb layers is roughly 2.52 nm, or 15.7\% of the interaction strength.  Thus, we hypothesize that the secondary transition we observe at low temperatures is indicative of the weak RKKY-like exchange that is still present, albeit significantly weaker than the bulk GdAuSb films. Therefore, through careful superlattice design and growth we are able to successfully observe a suppression in global magnetic ordering. As a result, We believe this platform offers promise in studying unique magnetic Heusler alloy superlattices, though more work is to be done regarding the specifics on this new phase.

In conclusion, we demonstrated here the epitaxial stabilization of GdAuSb films and GdAuSb/LaAuSb superlattices. The superlattices display sharp interfaces, as observed by strong fringes in X-ray diffraction and real space STEM imaging. Control of these interfaces between magnetically ordered GdAuSb and paramagnetic LaAuSb layers, combined with the bulk Dirac dispersion, provides the essential ingredients to explore the interplay of magnetism and topology.

\section{Methods and Materials}
\textbf{Thin film growth.} \textit{Ln}AuSb (\textit{Ln} = Gd, La) films and GdAuSb/LaAuSb superlattices were grown in a custom MBE system (MANTIS Deposition) on Al$_2$O$_3$ (0001) substrates (MTI) at a growth temperature of $650^\circ$C as measured by pyrometer. The lattice mismatch between GdAuSb (LaAuSb) and Al$_2$O$_3$ is estimated to be 4.9\% (2.6\%) tensile. Following the film growth, samples are capped with amorphous Ge or polycrystalline Sb to prevent oxidation upon removal from the vacuum system. La, Gd, and Au fluxes of $1.5\times10^{13}$ atoms/cm$^{2}$s were supplied from effusion cells, as measured by an \textit{in-situ} quartz crystal monitor (QCM). Due to the high relative volatility of Sb, a 30\% excess flux of Sb ($\sim2 \times 10^{13}$ atoms/cm$^2$s) was supplied using a cracker cell, similar to the strategy used for cubic Heusler compounds \cite{patel2014surface, kawasaki2014growth, kawasaki2018simple}. Absolute fluxes were calibrated \textit{ex-situ} by Rutherford Backscattering Spectrometry (RBS).

\textbf{X-ray Diffraction.} Phase purity and crystalline quality were confirmed \textit{ex-situ} by x-ray diffraction (XRD) using a Panalytical Empyrean diffractometer with Cu $K \alpha$ radiation. All scans used a quadruple-bounce Ge 220 monochromator on the incident beam. A second triple-bounce monochromator inserted in the diffracted beam path was inserted for the rocking curve measurements. 

\textbf{Photoemission Spectroscopy and band structure calculations.} Angle resolved and angle-integrated photoemission spectroscopy measurements were performed at beamline 29-ID of the Advanced Photon Source (APS), using a Scienta R4000 analyzer (angular acceptance angle of 14 degrees) and incident photon energies in the range 300 to 2000 eV. The Fermi level was determined via reference measurements on a gold screw that is in electrical contact with the sample. To protect the surfaces, these films were capped with Ge before removal from the MBE system, transferred through air, and then and then cleaved \textit{in-situ} after reloading into the ARPES endstation to expose a clean film surface. For comparison with experiment, density functional theory calculations were performed using the Perdew - Becke - Ernzerhof (PBE) parametrization of the generalized gradient approximation including fully relativistic spin-orbit coupling effects (GGA+SO), as implemented in WIEN2k \cite{wien2k}.

\textbf{Transmission Electron Microscopy.} The TEM specimen preparation began with coating the sample surface with platinum using a sputter coater (Leica ACE600) at room temperature. After that, we used the conventional lift-out methods using a focused ion beam (FIB, FEI Helios PFIB G4) with plasma beam energies of 30 keV for milling and 2 keV for the final FIB polishing step. For cleaning the TEM specimen, we used low-energy Ar-ion milling (Fischione Nanomill) for 25 minutes at 900 and 500 eV in tilting angles of $\pm$10$^{\circ}$. The TEM specimen was cleaned using a plasma cleaner (Ibss GV10x DS Asher plasma cleaner) for 5 min under 20W to remove hydrocarbon contamination before inserting into the TEM column. The GaAuSb/LaAuSb superlattice sample was characterized using aberration-corrected scanning transmission electron microscopy (STEM) in 200 keV low-angle annular dark-field imaging mode.

\textbf{Magnetotransport.} Magnetotransport properties were measured using a Quantum Design Physical Property Measurement System (PPMS). Measurements were taken in a standard Van der Pauw geometry using annealed In contacts at the corners of the sample.

\textbf{Superconducting Quantum Interference (SQUID) Magnetometry.} Magnetometry data was collected using a Quantum Design Magnetic Property Measurement System (MPMS) with a base tempearture of 1.5~K. Field cooled measurements were conducted at a field value of 0.4~T aligned with the $c$-axis of the crystal. For background subtraction, a bare Al$_2$O$_3$ (0001) substrate capped with amorphous germanium was measured under the same field cooled and zero-field cooled conditions as the sample and then this trace was subtracted from the film data to remove the background signal.

\begin{acknowledgments}
Synthesis was primarily supported by the DOE Office of Science, Synthesis and Processing Science (DE-SC0023958, by P.S. and J.K.K.). Diffraction, STEM imaging, transport, and magnetometry were supported by the MRSEC program of the National Science Foundation (DMR-2309000, P.S., S.I., Z.L., and T.S.). Preliminary synthesis was supported by the Army Research Office (W911NF-17-1-0254, by P.S., D. D., and E.S.). Transmission electron microscopy imaging was supported by the US Department of Energy (DE-SC0020313, by S.I. and P.M.V.). J.K.K. acknowledges support from the Gordon and Betty Moore Foundation, grant DOI 10.37807/GBMF13808, for revisions of this manuscript.

The authors gratefully acknowledge the use of x-ray diffraction and TEM instrumentation in the Wisconsin Center for Nanoscale Technology. This Center is partially supported by the Wisconsin Materials Research Science and Engineering Center (NSF DMR-2309000) and by the University of Wisconsin–Madison. This research used resources of the Advanced Photon Source, a U.S. Department of Energy (DOE) Office of Science User Facility operated for the DOE Office of Science by Argonne National Laboratory under Contract No. DE-AC02-06CH11357.

\end{acknowledgments}

\bibliography{bibliography}

\begin{thebibliography}{20}%
\makeatletter
\providecommand \@ifxundefined [1]{%
 \@ifx{#1\undefined}
}%
\providecommand \@ifnum [1]{%
 \ifnum #1\expandafter \@firstoftwo
 \else \expandafter \@secondoftwo
 \fi
}%
\providecommand \@ifx [1]{%
 \ifx #1\expandafter \@firstoftwo
 \else \expandafter \@secondoftwo
 \fi
}%
\providecommand \natexlab [1]{#1}%
\providecommand \enquote  [1]{``#1''}%
\providecommand \bibnamefont  [1]{#1}%
\providecommand \bibfnamefont [1]{#1}%
\providecommand \citenamefont [1]{#1}%
\providecommand \href@noop [0]{\@secondoftwo}%
\providecommand \href [0]{\begingroup \@sanitize@url \@href}%
\providecommand \@href[1]{\@@startlink{#1}\@@href}%
\providecommand \@@href[1]{\endgroup#1\@@endlink}%
\providecommand \@sanitize@url [0]{\catcode `\\12\catcode `\$12\catcode
  `\&12\catcode `\#12\catcode `\^12\catcode `\_12\catcode `\%12\relax}%
\providecommand \@@startlink[1]{}%
\providecommand \@@endlink[0]{}%
\providecommand \url  [0]{\begingroup\@sanitize@url \@url }%
\providecommand \@url [1]{\endgroup\@href {#1}{\urlprefix }}%
\providecommand \urlprefix  [0]{URL }%
\providecommand \Eprint [0]{\href }%
\providecommand \doibase [0]{https://doi.org/}%
\providecommand \selectlanguage [0]{\@gobble}%
\providecommand \bibinfo  [0]{\@secondoftwo}%
\providecommand \bibfield  [0]{\@secondoftwo}%
\providecommand \translation [1]{[#1]}%
\providecommand \BibitemOpen [0]{}%
\providecommand \bibitemStop [0]{}%
\providecommand \bibitemNoStop [0]{.\EOS\space}%
\providecommand \EOS [0]{\spacefactor3000\relax}%
\providecommand \BibitemShut  [1]{\csname bibitem#1\endcsname}%
\let\auto@bib@innerbib\@empty
\bibitem [{\citenamefont {Du}\ \emph {et~al.}(2019{\natexlab{a}})\citenamefont
  {Du}, \citenamefont {Lim}, \citenamefont {Zhang}, \citenamefont {Strohbeen},
  \citenamefont {Shourov}, \citenamefont {Rodolakis}, \citenamefont
  {McChesney}, \citenamefont {Voyles}, \citenamefont {Fredrickson},\ and\
  \citenamefont {Kawasaki}}]{du2019laglps}%
  \BibitemOpen
  \bibfield  {author} {\bibinfo {author} {\bibfnamefont {D.}~\bibnamefont
  {Du}}, \bibinfo {author} {\bibfnamefont {A.}~\bibnamefont {Lim}}, \bibinfo
  {author} {\bibfnamefont {C.}~\bibnamefont {Zhang}}, \bibinfo {author}
  {\bibfnamefont {P.~J.}\ \bibnamefont {Strohbeen}}, \bibinfo {author}
  {\bibfnamefont {E.~H.}\ \bibnamefont {Shourov}}, \bibinfo {author}
  {\bibfnamefont {F.}~\bibnamefont {Rodolakis}}, \bibinfo {author}
  {\bibfnamefont {J.~L.}\ \bibnamefont {McChesney}}, \bibinfo {author}
  {\bibfnamefont {P.}~\bibnamefont {Voyles}}, \bibinfo {author} {\bibfnamefont
  {D.~C.}\ \bibnamefont {Fredrickson}},\ and\ \bibinfo {author} {\bibfnamefont
  {J.~K.}\ \bibnamefont {Kawasaki}},\ }\bibfield  {title} {\bibinfo {title}
  {High electrical conductivity in the epitaxial polar metals {L}a{A}u{G}e and
  {L}a{P}t{S}b},\ }\href@noop {} {\bibfield  {journal} {\bibinfo  {journal}
  {APL Mater.}\ }\textbf {\bibinfo {volume} {7}},\ \bibinfo {pages} {121107}
  (\bibinfo {year} {2019}{\natexlab{a}})}\BibitemShut {NoStop}%
\bibitem [{\citenamefont {Du}\ \emph {et~al.}(2024)\citenamefont {Du},
  \citenamefont {Zhang}, \citenamefont {Wei}, \citenamefont {Teng},
  \citenamefont {Genser}, \citenamefont {Voyles}, \citenamefont {Rabe},\ and\
  \citenamefont {Kawasaki}}]{du2024polardist}%
  \BibitemOpen
  \bibfield  {author} {\bibinfo {author} {\bibfnamefont {D.}~\bibnamefont
  {Du}}, \bibinfo {author} {\bibfnamefont {C.}~\bibnamefont {Zhang}}, \bibinfo
  {author} {\bibfnamefont {J.}~\bibnamefont {Wei}}, \bibinfo {author}
  {\bibfnamefont {Y.}~\bibnamefont {Teng}}, \bibinfo {author} {\bibfnamefont
  {K.~T.}\ \bibnamefont {Genser}}, \bibinfo {author} {\bibfnamefont {P.~M.}\
  \bibnamefont {Voyles}}, \bibinfo {author} {\bibfnamefont {K.~M.}\
  \bibnamefont {Rabe}},\ and\ \bibinfo {author} {\bibfnamefont {J.~K.}\
  \bibnamefont {Kawasaki}},\ }\bibfield  {title} {\bibinfo {title} {Tunable
  polar distortions and magnetism in {G}d$_{x}${L}a$_{1-x}${P}t{S}b epitaxial
  films},\ }\href@noop {} {\bibfield  {journal} {\bibinfo  {journal} {Phys.
  Rev. Mater.}\ }\textbf {\bibinfo {volume} {8}},\ \bibinfo {pages} {104413}
  (\bibinfo {year} {2024})}\BibitemShut {NoStop}%
\bibitem [{\citenamefont {Du}\ \emph {et~al.}(2023)\citenamefont {Du},
  \citenamefont {Thoutam}, \citenamefont {Genser}, \citenamefont {Zhang},
  \citenamefont {Rabe}, \citenamefont {Samanta}, \citenamefont {Jung},
  \citenamefont {Jalan}, \citenamefont {voyles},\ and\ \citenamefont
  {Kawasaki}}]{du2023gdptsbvac}%
  \BibitemOpen
  \bibfield  {author} {\bibinfo {author} {\bibfnamefont {D.}~\bibnamefont
  {Du}}, \bibinfo {author} {\bibfnamefont {L.~R.}\ \bibnamefont {Thoutam}},
  \bibinfo {author} {\bibfnamefont {K.~T.}\ \bibnamefont {Genser}}, \bibinfo
  {author} {\bibfnamefont {C.}~\bibnamefont {Zhang}}, \bibinfo {author}
  {\bibfnamefont {K.~M.}\ \bibnamefont {Rabe}}, \bibinfo {author}
  {\bibfnamefont {T.}~\bibnamefont {Samanta}}, \bibinfo {author} {\bibfnamefont
  {T.}~\bibnamefont {Jung}}, \bibinfo {author} {\bibfnamefont {B.}~\bibnamefont
  {Jalan}}, \bibinfo {author} {\bibfnamefont {P.~M.}\ \bibnamefont {voyles}},\
  and\ \bibinfo {author} {\bibfnamefont {J.~K.}\ \bibnamefont {Kawasaki}},\
  }\bibfield  {title} {\bibinfo {title} {Effect of {P}t vacancies on
  magnetotransport of weyl semimetal candidate {G}d{P}t{S}b epitaxial films},\
  }\href@noop {} {\bibfield  {journal} {\bibinfo  {journal} {Phys. Rev.
  Materials}\ }\textbf {\bibinfo {volume} {7}},\ \bibinfo {pages} {084204}
  (\bibinfo {year} {2023})}\BibitemShut {NoStop}%
\bibitem [{\citenamefont {LaDuca}\ \emph {et~al.}(2024)\citenamefont {LaDuca},
  \citenamefont {Samanta}, \citenamefont {Hagopian}, \citenamefont {Jung},
  \citenamefont {Su}, \citenamefont {Genser}, \citenamefont {Rabe},
  \citenamefont {Voyles}, \citenamefont {Arnold},\ and\ \citenamefont
  {Kawasaki}}]{laduca2024flexo}%
  \BibitemOpen
  \bibfield  {author} {\bibinfo {author} {\bibfnamefont {Z.}~\bibnamefont
  {LaDuca}}, \bibinfo {author} {\bibfnamefont {T.}~\bibnamefont {Samanta}},
  \bibinfo {author} {\bibfnamefont {N.}~\bibnamefont {Hagopian}}, \bibinfo
  {author} {\bibfnamefont {T.}~\bibnamefont {Jung}}, \bibinfo {author}
  {\bibfnamefont {K.}~\bibnamefont {Su}}, \bibinfo {author} {\bibfnamefont
  {K.}~\bibnamefont {Genser}}, \bibinfo {author} {\bibfnamefont {K.~M.}\
  \bibnamefont {Rabe}}, \bibinfo {author} {\bibfnamefont {P.~M.}\ \bibnamefont
  {Voyles}}, \bibinfo {author} {\bibfnamefont {M.~S.}\ \bibnamefont {Arnold}},\
  and\ \bibinfo {author} {\bibfnamefont {J.~K.}\ \bibnamefont {Kawasaki}},\
  }\bibfield  {title} {\bibinfo {title} {Cold {S}eeded {E}pitaxy and
  {F}lexomagnetism in {G}d{A}u{G}e {M}embranes {E}xfoliated {F}rom
  {G}raphene/{G}e(111)},\ }\href@noop {} {\bibfield  {journal} {\bibinfo
  {journal} {Nano Lett.}\ }\textbf {\bibinfo {volume} {24}},\ \bibinfo {pages}
  {10284} (\bibinfo {year} {2024})}\BibitemShut {NoStop}%
\bibitem [{\citenamefont {Manna}\ \emph {et~al.}(2018)\citenamefont {Manna},
  \citenamefont {Sun}, \citenamefont {Muechler}, \citenamefont {K\"{u}bler},\
  and\ \citenamefont {Felser}}]{manna2018hwb}%
  \BibitemOpen
  \bibfield  {author} {\bibinfo {author} {\bibfnamefont {K.}~\bibnamefont
  {Manna}}, \bibinfo {author} {\bibfnamefont {Y.}~\bibnamefont {Sun}}, \bibinfo
  {author} {\bibfnamefont {L.}~\bibnamefont {Muechler}}, \bibinfo {author}
  {\bibfnamefont {J.}~\bibnamefont {K\"{u}bler}},\ and\ \bibinfo {author}
  {\bibfnamefont {C.}~\bibnamefont {Felser}},\ }\bibfield  {title} {\bibinfo
  {title} {Heusler, weyl and berry},\ }\href@noop {} {\bibfield  {journal}
  {\bibinfo  {journal} {Nat. Rev. Mater.}\ }\textbf {\bibinfo {volume} {3}},\
  \bibinfo {pages} {244} (\bibinfo {year} {2018})}\BibitemShut {NoStop}%
\bibitem [{\citenamefont {Duan}\ \emph {et~al.}(2018)\citenamefont {Duan},
  \citenamefont {Wu}, \citenamefont {Chen}, \citenamefont {Zhang},
  \citenamefont {Liu}, \citenamefont {Yuan},\ and\ \citenamefont
  {Cao}}]{duan2018repntuning}%
  \BibitemOpen
  \bibfield  {author} {\bibinfo {author} {\bibfnamefont {X.}~\bibnamefont
  {Duan}}, \bibinfo {author} {\bibfnamefont {F.}~\bibnamefont {Wu}}, \bibinfo
  {author} {\bibfnamefont {J.}~\bibnamefont {Chen}}, \bibinfo {author}
  {\bibfnamefont {P.}~\bibnamefont {Zhang}}, \bibinfo {author} {\bibfnamefont
  {Y.}~\bibnamefont {Liu}}, \bibinfo {author} {\bibfnamefont {H.}~\bibnamefont
  {Yuan}},\ and\ \bibinfo {author} {\bibfnamefont {C.}~\bibnamefont {Cao}},\
  }\bibfield  {title} {\bibinfo {title} {Tunable electronic structure and
  topological properties of {L}n{P}n ({L}n={C}e, {P}r, {S}m, {G}d, {Y}b;
  {P}n={S}b, {B}i)},\ }\href@noop {} {\bibfield  {journal} {\bibinfo  {journal}
  {Commun. Phys.}\ }\textbf {\bibinfo {volume} {1}},\ \bibinfo {pages} {71}
  (\bibinfo {year} {2018})}\BibitemShut {NoStop}%
\bibitem [{\citenamefont {Ho}\ \emph {et~al.}(2023)\citenamefont {Ho},
  \citenamefont {Hu}, \citenamefont {Quang~To}, \citenamefont {Bryant},\ and\
  \citenamefont {Janotti}}]{ho2023repntopology}%
  \BibitemOpen
  \bibfield  {author} {\bibinfo {author} {\bibfnamefont {D.~Q.}\ \bibnamefont
  {Ho}}, \bibinfo {author} {\bibfnamefont {R.}~\bibnamefont {Hu}}, \bibinfo
  {author} {\bibfnamefont {D.}~\bibnamefont {Quang~To}}, \bibinfo {author}
  {\bibfnamefont {G.~W.}\ \bibnamefont {Bryant}},\ and\ \bibinfo {author}
  {\bibfnamefont {A.}~\bibnamefont {Janotti}},\ }\bibfield  {title} {\bibinfo
  {title} {Emerging nontrivial topology in ultrathin films of rare-earth
  pnictides},\ }\href@noop {} {\bibfield  {journal} {\bibinfo  {journal} {ACS
  Nano}\ }\textbf {\bibinfo {volume} {17}},\ \bibinfo {pages} {20991} (\bibinfo
  {year} {2023})}\BibitemShut {NoStop}%
\bibitem [{\citenamefont {Kushnirenko}\ \emph {et~al.}(2022)\citenamefont
  {Kushnirenko}, \citenamefont {Schrunk}, \citenamefont {Kuthanazhi},
  \citenamefont {Wang}, \citenamefont {Ahn}, \citenamefont {O'Leary},
  \citenamefont {Eaton}, \citenamefont {Bud'ko}, \citenamefont {Slager},
  \citenamefont {Canfield},\ and\ \citenamefont
  {Kaminski}}]{kushnirenko2022repnfermi}%
  \BibitemOpen
  \bibfield  {author} {\bibinfo {author} {\bibfnamefont {Y.}~\bibnamefont
  {Kushnirenko}}, \bibinfo {author} {\bibfnamefont {B.}~\bibnamefont
  {Schrunk}}, \bibinfo {author} {\bibfnamefont {B.}~\bibnamefont {Kuthanazhi}},
  \bibinfo {author} {\bibfnamefont {L.-L.}\ \bibnamefont {Wang}}, \bibinfo
  {author} {\bibfnamefont {J.}~\bibnamefont {Ahn}}, \bibinfo {author}
  {\bibfnamefont {E.}~\bibnamefont {O'Leary}}, \bibinfo {author} {\bibfnamefont
  {A.}~\bibnamefont {Eaton}}, \bibinfo {author} {\bibfnamefont {S.~L.}\
  \bibnamefont {Bud'ko}}, \bibinfo {author} {\bibfnamefont {R.-J.}\
  \bibnamefont {Slager}}, \bibinfo {author} {\bibfnamefont {P.~C.}\
  \bibnamefont {Canfield}},\ and\ \bibinfo {author} {\bibfnamefont
  {A.}~\bibnamefont {Kaminski}},\ }\bibfield  {title} {\bibinfo {title}
  {Rare-earth monopnictides: Family of antiferromagnets hosting magnetic fermi
  arcs},\ }\href@noop {} {\bibfield  {journal} {\bibinfo  {journal} {Phys. Rev.
  B}\ }\textbf {\bibinfo {volume} {106}},\ \bibinfo {pages} {115112} (\bibinfo
  {year} {2022})}\BibitemShut {NoStop}%
\bibitem [{\citenamefont {Du}\ \emph {et~al.}(2019{\natexlab{b}})\citenamefont
  {Du}, \citenamefont {Lim}, \citenamefont {Zhang}, \citenamefont {Strohbeen},
  \citenamefont {Shourov}, \citenamefont {Rodolakis}, \citenamefont
  {McChesney}, \citenamefont {Voyles}, \citenamefont {Fredrickson},\ and\
  \citenamefont {Kawasaki}}]{ddu2019polarmetals}%
  \BibitemOpen
  \bibfield  {author} {\bibinfo {author} {\bibfnamefont {D.}~\bibnamefont
  {Du}}, \bibinfo {author} {\bibfnamefont {A.}~\bibnamefont {Lim}}, \bibinfo
  {author} {\bibfnamefont {C.}~\bibnamefont {Zhang}}, \bibinfo {author}
  {\bibfnamefont {P.~J.}\ \bibnamefont {Strohbeen}}, \bibinfo {author}
  {\bibfnamefont {E.~H.}\ \bibnamefont {Shourov}}, \bibinfo {author}
  {\bibfnamefont {F.}~\bibnamefont {Rodolakis}}, \bibinfo {author}
  {\bibfnamefont {J.~L.}\ \bibnamefont {McChesney}}, \bibinfo {author}
  {\bibfnamefont {P.}~\bibnamefont {Voyles}}, \bibinfo {author} {\bibfnamefont
  {D.~C.}\ \bibnamefont {Fredrickson}},\ and\ \bibinfo {author} {\bibfnamefont
  {J.~K.}\ \bibnamefont {Kawasaki}},\ }\bibfield  {title} {\bibinfo {title}
  {High electrical conductivity in the epitaxial polar metals laauge and
  laptsb},\ }\href@noop {} {\bibfield  {journal} {\bibinfo  {journal} {APL
  Mater.}\ }\textbf {\bibinfo {volume} {7}},\ \bibinfo {pages} {121107}
  (\bibinfo {year} {2019}{\natexlab{b}})}\BibitemShut {NoStop}%
\bibitem [{\citenamefont {Seibel}\ \emph {et~al.}(2014)\citenamefont {Seibel},
  \citenamefont {Schoop}, \citenamefont {Xie}, \citenamefont {Gibson},
  \citenamefont {Fuccillo}, \citenamefont {Krizan},\ and\ \citenamefont
  {Cava}}]{seibel2015gold}%
  \BibitemOpen
  \bibfield  {author} {\bibinfo {author} {\bibfnamefont {E.~M.}\ \bibnamefont
  {Seibel}}, \bibinfo {author} {\bibfnamefont {L.~M.}\ \bibnamefont {Schoop}},
  \bibinfo {author} {\bibfnamefont {W.}~\bibnamefont {Xie}}, \bibinfo {author}
  {\bibfnamefont {J.~B.}\ \bibnamefont {Gibson}, \bibfnamefont {Quinn
  D.~Webb}}, \bibinfo {author} {\bibfnamefont {M.~K.}\ \bibnamefont
  {Fuccillo}}, \bibinfo {author} {\bibfnamefont {J.~W.}\ \bibnamefont
  {Krizan}},\ and\ \bibinfo {author} {\bibfnamefont {R.~J.}\ \bibnamefont
  {Cava}},\ }\bibfield  {title} {\bibinfo {title} {Gold-gold bonding: The key
  to stabilizing the 19-electron ternary phases \textit{Ln}ausb
  (\textit{Ln}=la-nd and sm)},\ }\href@noop {} {\bibfield  {journal} {\bibinfo
  {journal} {J. Am. Chem. Soc.}\ }\textbf {\bibinfo {volume} {137}},\ \bibinfo
  {pages} {1282} (\bibinfo {year} {2014})}\BibitemShut {NoStop}%
\bibitem [{\citenamefont {Strohbeen}\ \emph {et~al.}(2019)\citenamefont
  {Strohbeen}, \citenamefont {Du}, \citenamefont {Zhang}, \citenamefont
  {Shourov}, \citenamefont {Rodolakis}, \citenamefont {McChesney},
  \citenamefont {Voyles},\ and\ \citenamefont
  {Kawasaki}}]{strohbeen2019LaAuSb}%
  \BibitemOpen
  \bibfield  {author} {\bibinfo {author} {\bibfnamefont {P.~J.}\ \bibnamefont
  {Strohbeen}}, \bibinfo {author} {\bibfnamefont {D.}~\bibnamefont {Du}},
  \bibinfo {author} {\bibfnamefont {C.}~\bibnamefont {Zhang}}, \bibinfo
  {author} {\bibfnamefont {E.~H.}\ \bibnamefont {Shourov}}, \bibinfo {author}
  {\bibfnamefont {F.}~\bibnamefont {Rodolakis}}, \bibinfo {author}
  {\bibfnamefont {J.~L.}\ \bibnamefont {McChesney}}, \bibinfo {author}
  {\bibfnamefont {P.~M.}\ \bibnamefont {Voyles}},\ and\ \bibinfo {author}
  {\bibfnamefont {J.~K.}\ \bibnamefont {Kawasaki}},\ }\bibfield  {title}
  {\bibinfo {title} {Electronically enhanced layer buckling and {A}u--{A}u
  dimerization in epitaxial {L}a{A}u{S}b films},\ }\href@noop {} {\bibfield
  {journal} {\bibinfo  {journal} {Phys. Rev. Materials}\ }\textbf {\bibinfo
  {volume} {3}},\ \bibinfo {pages} {024201} (\bibinfo {year}
  {2019})}\BibitemShut {NoStop}%
\bibitem [{\citenamefont {Brown-Heft}\ \emph {et~al.}(2018)\citenamefont
  {Brown-Heft}, \citenamefont {Logan}, \citenamefont {McFadden}, \citenamefont
  {Guillemard}, \citenamefont {Le~F\`{e}vre}, \citenamefont {Bertran},
  \citenamefont {Andrieu},\ and\ \citenamefont
  {Palmstr{\o}m}}]{brownheft2018heuslersl}%
  \BibitemOpen
  \bibfield  {author} {\bibinfo {author} {\bibfnamefont {T.~L.}\ \bibnamefont
  {Brown-Heft}}, \bibinfo {author} {\bibfnamefont {J.~A.}\ \bibnamefont
  {Logan}}, \bibinfo {author} {\bibfnamefont {A.~P.}\ \bibnamefont {McFadden}},
  \bibinfo {author} {\bibfnamefont {C.}~\bibnamefont {Guillemard}}, \bibinfo
  {author} {\bibfnamefont {P.}~\bibnamefont {Le~F\`{e}vre}}, \bibinfo {author}
  {\bibfnamefont {F.}~\bibnamefont {Bertran}}, \bibinfo {author} {\bibfnamefont
  {S.}~\bibnamefont {Andrieu}},\ and\ \bibinfo {author} {\bibfnamefont {C.~J.}\
  \bibnamefont {Palmstr{\o}m}},\ }\bibfield  {title} {\bibinfo {title}
  {Epitaxial {H}eusler superlattice {C}o$_{2}${M}n{A}l/{F}e$_{2}${M}n{A}l with
  perpendicular magnetic anisotropy and termination-dependent
  half-metallicity},\ }\href@noop {} {\bibfield  {journal} {\bibinfo  {journal}
  {Phys. Rev. Materials}\ }\textbf {\bibinfo {volume} {2}},\ \bibinfo {pages}
  {034402} (\bibinfo {year} {2018})}\BibitemShut {NoStop}%
\bibitem [{\citenamefont {Blaha}\ \emph {et~al.}(2018)\citenamefont {Blaha},
  \citenamefont {Schwarz}, \citenamefont {Madsen}, \citenamefont {Kvasnicka},
  \citenamefont {Luitz}, \citenamefont {Laskowski}, \citenamefont {Tran},\ and\
  \citenamefont {Marks}}]{wien2k}%
  \BibitemOpen
  \bibfield  {author} {\bibinfo {author} {\bibfnamefont {P.}~\bibnamefont
  {Blaha}}, \bibinfo {author} {\bibfnamefont {K.}~\bibnamefont {Schwarz}},
  \bibinfo {author} {\bibfnamefont {G.}~\bibnamefont {Madsen}}, \bibinfo
  {author} {\bibfnamefont {D.}~\bibnamefont {Kvasnicka}}, \bibinfo {author}
  {\bibfnamefont {J.}~\bibnamefont {Luitz}}, \bibinfo {author} {\bibfnamefont
  {R.}~\bibnamefont {Laskowski}}, \bibinfo {author} {\bibfnamefont
  {F.}~\bibnamefont {Tran}},\ and\ \bibinfo {author} {\bibfnamefont
  {L.}~\bibnamefont {Marks}},\ }\href@noop {} {\bibfield  {journal} {\bibinfo
  {journal} {\textit{WIEN2k}, An augmented plane wave + local orbitals program
  for calculated crystal properties}\ } (\bibinfo {year} {2018})}\BibitemShut
  {NoStop}%
\bibitem [{\citenamefont {Williams}(2009)}]{lbldatabooklet}%
  \BibitemOpen
  \bibfield  {author} {\bibinfo {author} {\bibfnamefont {G.~P.}\ \bibnamefont
  {Williams}},\ }\href@noop {} {\bibinfo {title} {X-ray data booklet, section
  1.1: Electron binding energies}} (\bibinfo {year} {2009})\BibitemShut
  {NoStop}%
\bibitem [{\citenamefont {Tanuma}\ \emph {et~al.}(2003)\citenamefont {Tanuma},
  \citenamefont {Powell},\ and\ \citenamefont {Penn}}]{tanuma2003calculation}%
  \BibitemOpen
  \bibfield  {author} {\bibinfo {author} {\bibfnamefont {S.}~\bibnamefont
  {Tanuma}}, \bibinfo {author} {\bibfnamefont {C.~J.}\ \bibnamefont {Powell}},\
  and\ \bibinfo {author} {\bibfnamefont {D.~R.}\ \bibnamefont {Penn}},\
  }\bibfield  {title} {\bibinfo {title} {Calculation of electron inelastic mean
  free paths (imfps) vii. reliability of the tpp-2m imfp predictive equation},\
  }\href@noop {} {\bibfield  {journal} {\bibinfo  {journal} {Surf. Interface
  Anal.}\ }\textbf {\bibinfo {volume} {35}},\ \bibinfo {pages} {268} (\bibinfo
  {year} {2003})}\BibitemShut {NoStop}%
\bibitem [{\citenamefont {Mancoff}\ \emph {et~al.}(1999)\citenamefont
  {Mancoff}, \citenamefont {Bobo}, \citenamefont {Richter}, \citenamefont
  {Bessho}, \citenamefont {Johnson}, \citenamefont {Sinclair}, \citenamefont
  {Nix}, \citenamefont {White},\ and\ \citenamefont
  {Clemens}}]{mancoff1999nimnsbc1bsl}%
  \BibitemOpen
  \bibfield  {author} {\bibinfo {author} {\bibfnamefont {F.~B.}\ \bibnamefont
  {Mancoff}}, \bibinfo {author} {\bibfnamefont {J.~F.}\ \bibnamefont {Bobo}},
  \bibinfo {author} {\bibfnamefont {O.~E.}\ \bibnamefont {Richter}}, \bibinfo
  {author} {\bibfnamefont {K.}~\bibnamefont {Bessho}}, \bibinfo {author}
  {\bibfnamefont {P.~R.}\ \bibnamefont {Johnson}}, \bibinfo {author}
  {\bibfnamefont {R.}~\bibnamefont {Sinclair}}, \bibinfo {author}
  {\bibfnamefont {W.~D.}\ \bibnamefont {Nix}}, \bibinfo {author} {\bibfnamefont
  {R.~L.}\ \bibnamefont {White}},\ and\ \bibinfo {author} {\bibfnamefont
  {B.~M.}\ \bibnamefont {Clemens}},\ }\bibfield  {title} {\bibinfo {title}
  {Growth and characterization of epitaxial {N}i{M}n{S}b/{P}t{M}n{S}b
  {C}1$_{b}$ {H}eusler alloy superlattices},\ }\href@noop {} {\bibfield
  {journal} {\bibinfo  {journal} {J. Mater. Res.}\ }\textbf {\bibinfo {volume}
  {14}},\ \bibinfo {pages} {1560} (\bibinfo {year} {1999})}\BibitemShut
  {NoStop}%
\bibitem [{\citenamefont {Roth}\ \emph {et~al.}(1966)\citenamefont {Roth},
  \citenamefont {Zeiger},\ and\ \citenamefont {Kaplan}}]{roth1966rkky}%
  \BibitemOpen
  \bibfield  {author} {\bibinfo {author} {\bibfnamefont {L.~M.}\ \bibnamefont
  {Roth}}, \bibinfo {author} {\bibfnamefont {H.~J.}\ \bibnamefont {Zeiger}},\
  and\ \bibinfo {author} {\bibfnamefont {T.~A.}\ \bibnamefont {Kaplan}},\
  }\bibfield  {title} {\bibinfo {title} {Generalization of the
  ruderman-kittel-kasuya-yosida interaction for nonspherical fermi surfaces},\
  }\href@noop {} {\bibfield  {journal} {\bibinfo  {journal} {Phys. Rev.}\
  }\textbf {\bibinfo {volume} {149}},\ \bibinfo {pages} {519} (\bibinfo {year}
  {1966})}\BibitemShut {NoStop}%
\bibitem [{\citenamefont {Patel}\ \emph {et~al.}(2014)\citenamefont {Patel},
  \citenamefont {Kawasaki}, \citenamefont {Logan}, \citenamefont {Schultz},
  \citenamefont {Adell}, \citenamefont {Thiagarajan}, \citenamefont
  {Mikkelsen},\ and\ \citenamefont {Palmstr{\o}m}}]{patel2014surface}%
  \BibitemOpen
  \bibfield  {author} {\bibinfo {author} {\bibfnamefont {S.~J.}\ \bibnamefont
  {Patel}}, \bibinfo {author} {\bibfnamefont {J.~K.}\ \bibnamefont {Kawasaki}},
  \bibinfo {author} {\bibfnamefont {J.}~\bibnamefont {Logan}}, \bibinfo
  {author} {\bibfnamefont {B.~D.}\ \bibnamefont {Schultz}}, \bibinfo {author}
  {\bibfnamefont {J.}~\bibnamefont {Adell}}, \bibinfo {author} {\bibfnamefont
  {B.}~\bibnamefont {Thiagarajan}}, \bibinfo {author} {\bibfnamefont
  {A.}~\bibnamefont {Mikkelsen}},\ and\ \bibinfo {author} {\bibfnamefont
  {C.~J.}\ \bibnamefont {Palmstr{\o}m}},\ }\bibfield  {title} {\bibinfo {title}
  {Surface and electronic structure of epitaxial {P}t{L}u{S}b (001) thin
  films},\ }\href@noop {} {\bibfield  {journal} {\bibinfo  {journal} {Applied
  Physics Letters}\ }\textbf {\bibinfo {volume} {104}},\ \bibinfo {pages}
  {201603} (\bibinfo {year} {2014})}\BibitemShut {NoStop}%
\bibitem [{\citenamefont {Kawasaki}\ \emph {et~al.}(2014)\citenamefont
  {Kawasaki}, \citenamefont {Johansson}, \citenamefont {Schultz},\ and\
  \citenamefont {Palmstr{\o}m}}]{kawasaki2014growth}%
  \BibitemOpen
  \bibfield  {author} {\bibinfo {author} {\bibfnamefont {J.~K.}\ \bibnamefont
  {Kawasaki}}, \bibinfo {author} {\bibfnamefont {L.~I.}\ \bibnamefont
  {Johansson}}, \bibinfo {author} {\bibfnamefont {B.~D.}\ \bibnamefont
  {Schultz}},\ and\ \bibinfo {author} {\bibfnamefont {C.~J.}\ \bibnamefont
  {Palmstr{\o}m}},\ }\bibfield  {title} {\bibinfo {title} {Growth and transport
  properties of epitaxial lattice matched half heusler
  {C}o{T}i{S}b/{I}n{A}l{A}s/{I}n{P} (001) heterostructures},\ }\href@noop {}
  {\bibfield  {journal} {\bibinfo  {journal} {Applied Physics Letters}\
  }\textbf {\bibinfo {volume} {104}},\ \bibinfo {pages} {022109} (\bibinfo
  {year} {2014})}\BibitemShut {NoStop}%
\bibitem [{\citenamefont {Kawasaki}\ \emph {et~al.}(2018)\citenamefont
  {Kawasaki}, \citenamefont {Sharan}, \citenamefont {Johansson}, \citenamefont
  {Hjort}, \citenamefont {Timm}, \citenamefont {Thiagarajan}, \citenamefont
  {Schultz}, \citenamefont {Mikkelsen}, \citenamefont {Janotti},\ and\
  \citenamefont {Palmstr{\o}m}}]{kawasaki2018simple}%
  \BibitemOpen
  \bibfield  {author} {\bibinfo {author} {\bibfnamefont {J.~K.}\ \bibnamefont
  {Kawasaki}}, \bibinfo {author} {\bibfnamefont {A.}~\bibnamefont {Sharan}},
  \bibinfo {author} {\bibfnamefont {L.~I.}\ \bibnamefont {Johansson}}, \bibinfo
  {author} {\bibfnamefont {M.}~\bibnamefont {Hjort}}, \bibinfo {author}
  {\bibfnamefont {R.}~\bibnamefont {Timm}}, \bibinfo {author} {\bibfnamefont
  {B.}~\bibnamefont {Thiagarajan}}, \bibinfo {author} {\bibfnamefont {B.~D.}\
  \bibnamefont {Schultz}}, \bibinfo {author} {\bibfnamefont {A.}~\bibnamefont
  {Mikkelsen}}, \bibinfo {author} {\bibfnamefont {A.}~\bibnamefont {Janotti}},\
  and\ \bibinfo {author} {\bibfnamefont {C.~J.}\ \bibnamefont {Palmstr{\o}m}},\
  }\bibfield  {title} {\bibinfo {title} {A simple electron counting model for
  half-heusler surfaces},\ }\href@noop {} {\bibfield  {journal} {\bibinfo
  {journal} {Science advances}\ }\textbf {\bibinfo {volume} {4}},\ \bibinfo
  {pages} {eaar5832} (\bibinfo {year} {2018})}\BibitemShut {NoStop}%
\end{thebibliography}%

\end{document}